\documentclass[a4paper]{jpconf}

\usepackage{color}
\usepackage{amsmath}
\usepackage{graphicx}
\usepackage{dcolumn}
\usepackage{bm}
\usepackage{cancel}
\usepackage{soul}

\renewcommand{\vec}[1]{\mathbf{#1}}
\newcommand{\dint}[2]{\mathrm{d}^{#1}{#2}}

\begin{document}

\title{Finite elements and the discrete variable representation in nonequilibrium Green's function calculations. Atomic and molecular models}

\author{Karsten Balzer, Sebastian Bauch, and Michael Bonitz}

\address{Institut f\"ur Theoretische Physik und Astrophysik, Christian-Albrechts-Universit\"at Kiel, Leibnizstrasse 15, 24098 Kiel, Germany}

\ead{balzer@theo-physik.uni-kiel.de}

\begin{abstract}
In this contribution, we discuss the finite-element discrete variable representation~(FE-DVR) of the nonequilibrium Green's function and its implications on the description of strongly inhomogeneous quantum systems. In detail, we show that the complementary features of FEs and the DVR allows for a notably more efficient solution of the two-time Schwinger/Keldysh/Kadanoff-Baym equations compared to a general basis approach. Particularly, the use of the FE-DVR leads to an essential speedup in computing the self-energies.

As atomic and molecular examples we consider the He atom and the linear version of H$_3^+$ in one spatial dimension. For these closed-shell models we, in Hartree-Fock and second Born approximation, compute the ground-state properties and compare with the exact findings obtained from the solution of the few-particle time-dependent Schr\"odinger equation.
\end{abstract}

\section{Introduction}
In the last decade, the application of the nonequilibrium Green's function (NEGF) to describe strongly inhomogeneous quantum systems has started to become an actively considered subject. Thereby, various finite and localized systems have challenged attention and different many-body approximations have been applied. Recent state-of-the-art approaches discuss small atoms and molecules~\cite{stan06,dahlen07_prl}, few-electron quantum dots~\cite{balzer09_jpa, balzer09_prb} and quantum dots coupled to leads~\cite{myohanen08}, molecular junctions~\cite{thygesen08}, and Hubbard nanoclusters~\cite{vanfriesen09}. In part, these works also include the monitoring of the system's temporal evolution which, accounting for correlation and memory effects, requires to solve the two-time Schwinger/Keldysh/Kadanoff-Baym equations~\cite{martin59,keldysh64,kadanoff62} (SKKBE).

To solve the SKKBE for homogeneous quantum systems~\cite{danielewicz84,koehler95,bozek97,kwong00,bonitz96,semkat99,semkat00,binder97,kwong98,schaefer,banyai,jahnke} has become routine. However, this still does not hold for finite, localized and inhomogeneous systems. The reason is, that in contrast to homogeneous systems where one spatial coordinate or momentum drops out of the NEGF, the inhomogeneity claims adequate resolution in both coordinates or momenta. To meet these requirements all above mentioned works rely on an expansion of the NEGF in one-particle orbitals:~for atomic and molecular systems, linear combinations of Slater-type or Gauss-type orbitals are being used (also in connection with tight-binding models), whereas for other classes of problems e.g.~potential-eigenstate basis functions are being utilized. Beyond a basis ansatz, alternatives such as direct grid (finite-difference) methods do not exist as they are computationally very expensive. Nevertheless, also basis approaches are so far limited to relatively small basis sets. Generally, the numerical complexity involved in the description of the binary interactions does not permit to extend nonequilibrium calculations to much larger basis dimensions than a guide number of $n_b<50$ orbitals. This explains the low resolution in the description of photoionization processes of model atoms~\cite{hochstuhl09,bonitz09} using nonequilibrium Green's functions.

In this contribution, we---for the first time---develop a grid-based approach in the frame of the finite-element discrete variable representation (FE-DVR), e.g.~Refs.~\cite{rescigno00,collins04,schneider06,feist09} and references therein. This method leads to specially-designed, flexible basis sets which are capable to combine the advantages of pure grid and standard basis approaches, see Sec.~\ref{sec:fedvr} and also Ref.~\cite{balzer09_pra}. In particular, it allows for a very efficient treatment of the binary interactions and, in turn, a drastic simplification of higher-order self-energy expressions, which, generally, require the main effort in all NEGF calculations. As a result, only ${\cal O}(n_b^2)$ semi-analytical matrix elements of the interaction energy operator are required to compose the second Born self-energy. This has to be compared to a general basis representation of the NEGF: There, ${\cal O}(n_b^4)$ matrix elements are involved, cf.~\cite{balzer09_prb}. As a consequence, the use of the FE-DVR enables more efficient calculations at less storage memory and computing time and provides the basis to consider also spatially extended hamiltonians, where particles may occupy large domains in coordinate space.

After the outline of the method, we, in Sec.~\ref{sec:ammodels}, apply the FE-DVR representation of the NEGF to compute the ground-state properties of atomic and molecular models. First, in Sec.~\ref{subsec:he}, we discuss the one-dimensional (1D) helium atom~\cite{pindzola91,grobe93,bauer97,liu99,dahlen01,lein00,zanghellini04,ruggenthaler09}, where we focus on technical details such as grid size, DVR basis size, and convergence in the case of the Hartree-Fock and second Born approximation. The benchmarking results shown are also of relevance for time-dependent calculations, as they define and border the requirements to resolve explicit correlation effects (within a NEGF approach) e.g.~the two-electron resonances~\cite{tanner00} which are embedded in the one-electron continuum of the atom and are going to be occupied during laser-atom interactions---see also~\cite{david_pngf4} in the present volume. Finally, we consider the linear molecular ion H$_3^+$ in the symmetric 1D singlet configuration~\cite{kawata01,suzuki07} as an example of a two-electron molecule and vary the interatomic distance to record the binding-energy curves in Hartree-Fock and second Born approximation, see Sec.~\ref{subsec:h3+}. From this we can extract the minimum ground-state energies and the respective bond-lengths, which are compared to the findings from the few-particle time-dependent Schr\"odinger equation.

\section{\label{sec:fedvr}The nonequilibrium Green's function in FE-DVR representation}
For the description of the NEGF using finite elements together with the discrete variable representation~\cite{light85}~(DVR), we consider the general $N$-electron Hamiltonian
\begin{eqnarray}
\label{ham}
 \hat{h}=\sum_{i=1}^{N}\left(\hat{t}_i+\hat{v}_i\right)+\sum_{i<j}\hat{u}_{ij}\;,
\end{eqnarray}
with the kinetic energy $\hat{t}_i=-\nabla^2_i/2$, the time-dependent potential energy $\hat{v}_i=v(x_i,t)$, and the pair interaction energy $\hat{u}_{ij}=u(|x_i-x_j|)$  formulated in atomic units. The one-particle nonequilibrium Green's function with space-time arguments $1=(x,t)$ and $1'=(x',t')$ reads
\begin{eqnarray}
\label{gfdef}
 G(1,1')=-i\left\langle T_{\cal C}\psi(1)\psi^\dagger(1')\right\rangle\;,
\end{eqnarray}
where spin is omitted, $\psi(1)$ and $\psi^\dagger(1')$ are electron field operators, and $T_{\cal C}$ denotes time-ordering on the full Keldysh contour~\cite{keldysh64} $\cal C$. According to system (\ref{ham}), $G(1,1')$ obeys the SKKBE~\cite{martin59,keldysh64,kadanoff62}
\begin{eqnarray}
\label{kbe}
 \left\{\mathrm{i}\,\partial_{t}-H(1)\right\}\,G(1,1')=\delta_{\cal C}(1-1')\,+\,\int_{\cal C}\dint{}{2}\,\Sigma[G](1,2)\,G(2,1')\;,
\end{eqnarray}
with addition of its adjoint equation. Further, $H(1)$ is the one-electron (kinetic plus potential) energy, $\Sigma(1,1')$ denotes the self-energy, and equilibrium initial correlations are treat in the mixed Green's function approach~\cite{dahlen05,dahlen06,stan09}, cf.~Sec.~\ref{subsec:skkbe}.

The favorable aspects of the FE-DVR representation have been successfully used for the time-dependent Schr\"odinger equation (TDSE) by Rescigno~\cite{rescigno00} and others, e.g.~\cite{feist09}. There, the accuracy of the DVR~\cite{light07} on the one
hand and the sparse character of FEs on the other led to an efficient TDSE code which is well parallelizable. In our case, this hybrid approach allows us to rewrite the SKKBE~(\ref{kbe}) in a highly effective matrix notation using optimal combinations of a grid and a local basis.

\begin{figure}[t]
\begin{center}
 \includegraphics[width=35.0pc]{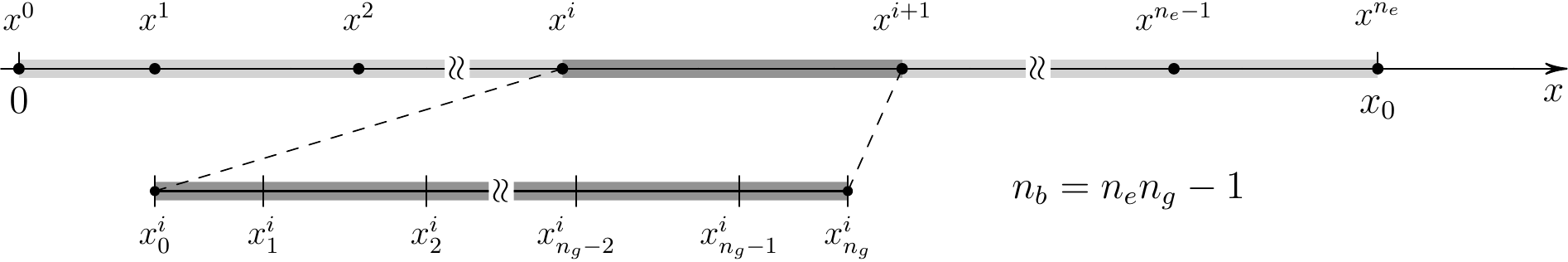}
\end{center}
 \caption{
Interval $[0,x_0]$, as discretized in FE-DVR representation with a number of $n_e$ finite elements and $n_g$ generalized Gauss-Lobatto points (per element). $n_b$ denotes the dimensionality of the extended basis covering the whole interval.
}\label{fig1}
\end{figure}

\subsection{\label{subsec:ansatz}Grid-based ansatz}
On a predefined spatial 1D interval ${\cal I}=[0,x_0]$, we expand the nonequilibrium Green's function of hamiltonian (\ref{ham}) as
\begin{eqnarray}
\label{gf}
 G(1,1')=\sum_{i_1m_1}\sum_{i_2m_2}\chi_{m_1}^{i_1}(x)\,\chi_{m_2}^{i_2}(x')\,g_{m_1m_2}^{i_1i_2}(t,t')\;,\hspace{2pc}x,x'\in {\cal I}\;,
\end{eqnarray}
with time-dependent complex coefficients $g_{mm'}^{ii'}(t,t')$ and real basis functions $\chi_m^i(x)$.
Outside the interval ${\cal I}$, we assume the NEGF  to vanish. All indices $i$ (superscripts), ranging $0,1,\ldots,n_e-1$ in Eq.~(\ref{gf}), are linked to a grid of length $x_0$ composed of $n_e$ finite elements, see Fig.~\ref{fig1}. All indices $m$ (subscripts), ranging $0,1,\ldots,n_g-1$, are connected to locally defined basis functions $\chi_m^i(x)$ which are constructed as follows: First, we divide the interval $\cal I$ into finite elements with boundaries $x^0=0<x^1<x^2<\ldots<x^{n_e-1},x^{n_e}=x_0$. To each FE $[x^i,x^{i+1}]$, we then attach a local DVR basis using the generalized Gauss-Lobatto (GGL) points~\cite{rescigno00} $x^i_m$ and weights $w_m^i$:
\begin{eqnarray}
\label{xim}
 x^i_m&=&\frac{1}{2}\left\{\left(x^{i+1}-x^i\right)x_m + \left(x^{i+1}+x^i\right)\right\}\;,\\
\label{wim}
 w^i_m&=&\frac{w_m}{2}\left(x^{i+1}-x^i\right)\;,
\end{eqnarray}
with the standard Gauss-Lobatto points $x_m$ (and weights $w_m$). For the special case of Legendre interpolating functions, the points $x_m$ are defined as roots of the first derivative of Legendre polynomials $P_n(x)$ according to
\begin{eqnarray}
\left.\frac{\mathrm{d}}{\mathrm{d}x}P_{n_g}(x)\right|_{x=x_m}=0\;,
\end{eqnarray}
and the weights $w_m$ are given by
\begin{eqnarray}
w_m=\displaystyle\frac{2}{n_g(n_g+1)[P_{n_g}(x_m)]^2}\;,
\end{eqnarray}
where $n_g$ denotes the total number of basis functions per element. In our approach, we keep the number of basis functions constant in each FE \cite{generalization}. However, in order to combine the locally defined DVR basis functions to a continuous basis set, the FE-DVR space is being spanned by two classes of functions---'bridge' and 'element' functions, see also Ref.~\cite{balzer09_pra}. The bridge function (the case $m=0$ in Eq.~(\ref{basis})) extends over two adjacent FEs and ensures \textit{communication} between the grid domains $i$ and $i+1$. In particular, it guarantees the continuity of the NEGF. The 'element' functions are zero at and outside the respective element boundaries. Using definitions (\ref{xim}) and (\ref{wim}), the basis functions have the explicit form
\begin{eqnarray}
\label{basis}
 \chi_{m}^{i}(x)&=
\left\{
\begin{array}{cc}
 \displaystyle\frac{f_{n_g-1}^{i}(x)\,+\,f_{0}^{i+1}(x)}{\sqrt{w^i_{n_g-1}+w^{i+1}_0}}&\;\;,\;m=0\hspace{1pc}\textup{('bridge' function)}\\
&\\
 \displaystyle\frac{f_{m}^{i}(x)}{\sqrt{w^i_m}} & ,\;\textup{else\hspace{1pc}('element' function)}
\end{array}
\right.\;,
\end{eqnarray}
and are orthonormal in the sense of the generalized Gauss-Lobatto quadrature. In Eq.~(\ref{basis}), the so-called Lobatto shape functions\cite{rescigno00,manolopoulos88} $f_{m}^{i}(x)$ are defined as
\begin{eqnarray}
f_{m}^{i}(x)&=&\left\{
\begin{array}{ccc}
\prod_{\bar{m}\neq m} \displaystyle\frac{x-x_{\bar{m}}^i}{x_m^i-x_{\bar{m}}^i} &\;\;,\;x^i\leq x\leq x^{i+1}\\
& \\
0&\hspace{4pc}\;\;,\;x<x^i\hspace{1pc}\textup{and}\hspace{1pc}x>x^{i+1}
\end{array}\right.\;,
\end{eqnarray}
and obey $f^i_m(x_{m'}^{i'})=\delta_{ii'}\delta_{mm'}$. As in the last finite element $n_e-1$ (i.e.~in the element $[x^{n_g-1},x_0]$) no bridge function is needed due to the boundary condition of vanishing $G(1,1')$ outside the interval $I$, the total FE-DVR set consists of
\begin{eqnarray}
\label{basisdim}
 n_b=n_e n_g-1\;.
\end{eqnarray}
basis functions. Finally, we note, that a generalization of Eq.~(\ref{basis}) to higher dimensions (2D and 3D) is, in principle, possible by using a product ansatz for the coordinate functions~\cite{schneider06}.

\subsection{\label{subsec:matrixelements}Matrix elements of the kinetic, potential and interaction energy operators}
With representation (\ref{gf}) of the NEGF, the SKKBE will transform into an equation of motion for the matrix $\vec{g}(t,t')$ with elements $g_{m_1m_2}^{i_1i_2}(t,t')$, cf.~Sec.~\ref{subsec:skkbe}. Obviously, this equation involves also the kinetic, potential and interaction energy operator of Eq.~(\ref{ham}) in matrix form, which we specify in the following. Thereby, integrations over coordinate space are performed by using the generalized Gauss-Lobatto quadrature rule, and case differentiations arise from the basis functions $\chi_m^i(x)$ being split into element and bridge functions.

First, let us consider the potential and the kinetic energy in FE-DVR representation: The potential-energy matrix is given by
\begin{eqnarray}
\label{vmat}
 v_{m_1m_2}^{i_1i_2}(t)=\int_{0}^{x_0}\!\!\!\dint{}{x}\,\chi_{m_1}^{i_1}(x)\,v(x,t)\,\chi_{m_2}^{i_2}(x)&\nonumber\\
                       =\delta_{i_1i_2}\,\delta_{m_1m_2}\,\tilde{v}_{m_1}^{i_1}(t)\;,
\end{eqnarray}
with
\begin{eqnarray}
 \tilde{v}_{m}^{i}(t)=
\left\{
\begin{array}{cc}
 v(x^{i}_{m},t) &,\;m>0\\
&\\
\displaystyle\frac{v(x_{n_g-1}^{i},t)\,w_{n_g-1}^{i}+v(x_{0}^{i+1},t)\,w_{0}^{i+1}}{w_{n_g-1}^{i}+w_{i+1}^{0}} &,\;m=0
\end{array}
\right.\;.
\end{eqnarray}
This implies that the potential energy is diagonal with respect to elements $i$ and local DVR basis indices $m$, and that, consequently, it can be represented by a vector of dimension $n_b$. Moreover, Eq.~(\ref{vmat}) holds true also for any other local operator. As the operator of the kinetic energy is non-local in coordinate space, the matrix elements $t_{m_1m_2}^{i_1i_2}$ are not diagonal. We follow the derivation of Ref.~\cite{rescigno00} and obtain the kinetic-energy matrix as
 \begin{eqnarray}
 t_{m_1m_2}^{i_1i_2}&=&-\frac{1}{2}\int_{0}^{x_0}\!\!\!\dint{}{x}\,\chi_{m_1}^{i_1}(x)\,\nabla^2\,\chi_{m_2}^{i_2}(x)\\
\label{tmat}
&=&\left\{
\begin{array}{cc}
 \frac{1}{2}\,\delta_{i_1i_2}\,\tilde{t}_{m_1m_2}^{\,i_1}\left[w_{m_1}^{i_1}w_{m_2}^{i_1}\right]^{-1/2}\; &\;,\; m_1>0,\,m_2>0\\
&\\
\frac{1}{2}\left(\delta_{i_1i_2}\,\tilde{t}_{n_g-1,m_2}^{\,i_1}+\delta_{i_1i_2-1}\,\tilde{t}_{0m_2}^{\,i_2}\right)\left[w_{n_g-1}^{i_1}+w_{0}^{i_1+1}\right]^{-1/2}\;&\;,\; m_1=0,\,m_2>0\\
&\\
\frac{1}{2}\left(\delta_{i_1i_2}\,\tilde{t}_{m_1n_g-1}^{\,i_1}+\delta_{i_1i_2+1}\,\tilde{t}_{m_10}^{\,i_1}\right)\left[w_{m_1}^{i_1}\left(w_{n_g-1}^{i_2}+w_{0}^{i_2+1}\right)\right]^{-1/2}\; &\;,\; m_1>0,\,m_2=0\\
&\\
\displaystyle\frac{\delta_{i_1i_2}\left(\tilde{t}_{n_g-1,n_g-1}^{\,i_1}+\tilde{t}_{00}^{\,i_1+1}\right)+\delta_{i_1i_2-1}\,\tilde{t}_{0, n_g-1}^{\,i_2}+\delta_{i_1i_2+1}\,\tilde{t}_{n_g-1,0}^{\,i_1}}{2\left[\left(w_{n_g-1}^{i_1}+w_{0}^{i_1+1}\right)\left(w_{n_g-1}^{i_2}+w_{0}^{i_2+1}\right)\right]^{1/2}}\; &,\; m_1=m_2=0
\end{array}
\right.\nonumber
\end{eqnarray}
Here, the case differentiations lead to the matrix having a block-diagonal form~\cite{schneider06}, and the quantity $\tilde{t}^{\,i}_{m_1m_2}$ is given by
\begin{eqnarray}
 \tilde{t}^{\,i}_{m_1m_2}=\sum_{m}\frac{\mathrm{d}  f_{m_1}^{i}(x^{i}_{m})}{\mathrm{d}x}\,\frac{\mathrm{d}  f_{m_2}^{i}(x^{i}_{m})}{\mathrm{d}x}\,w_{m}^{i}\;,
\end{eqnarray}
which involves the first derivatives of the Lobatto shape functions at the GGL points, see also Ref.~\cite{rescigno00}.

Next, let us focus on the matrix elements of the binary-interaction operator $\hat{u}$ (the two-electron integrals) which are carrying a set of four index-pairs $(i,m)$ and are defined by
\begin{eqnarray}
\label{umat}
 u_{m_1m_2,m_3m_4}^{i_1i_2,i_3i_4}&=&\int_{0}^{x_0}\!\!\!\dint{}{x}\!\int_{0}^{x_0}\!\!\!\dint{}{x'}\,\chi_{m_1}^{i_1}(x)\,\chi_{m_3}^{i_3}(x')\,u(|x-x'|)\,\chi_{m_2}^{i_2}(x)\,\chi_{m_4}^{i_4}(x')\;.
\end{eqnarray}
In a general basis approach, the two-electron integrals [Eq.~(\ref{umat}) with all index-pairs replaced by single indices] often require a careful analysis, as they are not analytically accessible and have to be numerically precomputed for all combination of indices, e.g.~\cite{balzer09_prb}. Although symmetry relations~\cite{symmetryrel} help to restrict oneself to a subset of indices, the effort scales with ${\cal O}(n_b^4)$ and thus can be huge for larger basis sets. On the contrary, using the FE-DVR, the evaluation of the two-electron integrals turns out to be much simpler. In particular, the integrals can be performed in a semi-analytical way such that Eq.~(\ref{umat}) reduces to
\begin{eqnarray}
\label{umat1}
 u_{m_1m_2,m_3m_4}^{i_1i_2,i_3i_4}&=&\delta_{i_1i_2}\delta_{i_3i_4}\delta_{m_1m_2}\delta_{m_3m_4}\tilde{u}_{m_1m_2}^{i_1i_2}\;.
\end{eqnarray}
where the kernel matrix $\tilde{\vec{u}}$ is symmetric and follows as
\begin{eqnarray}
\label{umatred}
 \tilde{u}_{m_1m_2}^{i_1i_2}=\sum_{i_3m_3} \alpha_{m_3}^{i_3} \beta^{i_1i_3}_{m_1m_3} \beta^{i_2i_3}_{m_2m_3}\;.
\end{eqnarray}
To obtain Eq.~(\ref{umatred}), we have used the separable form of the discretized interaction potential $u(|x-x'|)$, and, correspondingly, the quantities $\alpha_{m}^{i}$ denote the eigenvalues of the matrix
\begin{eqnarray}
\label{umatsep}
U_{(im)(i'm')}=u(|x_{m}^{i}-x_{m'}^{i'}|)=\sum_{i_3m_3}\alpha_{m_3}^{i_3}\tilde{\beta}_{i_3i}^{m_3m}\tilde{\beta}_{i_3i'}^{m_3m'}\;,
\end{eqnarray}
and $\beta_{mm'}^{ii'}$ are related to the eigenvectors $\tilde{\beta}_{mm'}^{ii'}$:
\begin{eqnarray}
 \beta_{mm'}^{ii'}=
\left\{
\begin{array}{cc}
\tilde{\beta}_{m'm}^{i'i} &,\;m>0\\
&\\
\displaystyle\frac{\tilde{\beta}_{m' (n_g-1)}^{i'i} w_{n_g-1}^{i}+\tilde{\beta}_{m' 0}^{i'(i+1)} w_{0}^{i+1}}{w_{n_g-1}^{i}+w_{i+1}^{0}} &,\;m=0
\end{array}
\right.\;.
\end{eqnarray}
In comparison with any single-electron matrix element (such as the kinetic or potential energy), in FE-DVR, the calculation of the binary-interaction matrix elements involves just an additional but numerically elementary matrix diagonalization. Furthermore, besides the fact that with Eq.~(\ref{umat1}) the two-electron integrals attain a very simple form independent of the specific pair-interaction potential, only a single  matrix of dimension $n_b\times n_b$ needs to be stored in the code. This memory-friendly property is based on the high degree of diagonality determined by the product of Kronecker deltas and represents a main attractive feature of the FE-DVR representation. Particularly, this aspect opens the way towards efficient NEGF calculations, since Eq.~(\ref{umat1}) has direct consequences for the structure of the self-energies, see the following Section.

\subsection{\label{subsec:skkbe}Equations of motion}
Once all relevant matrix elements are known with respect to the chosen FE-DVR basis (\ref{basis}), we can start to solve the Schwinger/Keldysh/Kadanoff-Baym equations (\ref{kbe}) for the one-particle Green's function $G(1,1')$ expanded in the form of Eq.~(\ref{gf}). This implies, though, the SKKBE in the matrix form of the finite-element discrete variable representation:
\begin{eqnarray}
\label{skkbe}
\sum_{im} \left\{\mathrm{i}\partial_t\delta_{m_1m}^{i_1i}-h_{m_1m}^{i_1i}(t)\right\}\, g_{mm_2}^{ii_2}(t,t')\!&=&\!\delta_{\cal C}(t-t')\,+\,\sum_{im}\int_{\cal C}\!\dint{}{t_2}\,\Sigma_{m_1m}^{i_1i}(t,t_2)\,g_{mm_2}^{ii_2}(t_2,t')\;,\hspace{1.5pc}\\
\label{skkbesigma}
\Sigma_{m_1m_2}^{i_1i_2}(t,t')\!&=&\!\delta_{\cal C}(t-t')\,\Sigma_{m_1m_2}^{\mathrm{HF},i_1i_2}(t)\,+\,\Sigma_{m_1m_2}^{\mathrm{corr},i_1i_2}(t,t')\;,
\end{eqnarray}
where we have denoted $\delta_{mm'}^{ii'}=\delta_{ii'}\delta_{mm'}$, 
$\vec{h}(t)=\vec{t}+\vec{v}(t)$, and
Eq.~(\ref{skkbe}) has to be supplied with its adjoint equation. Further, Eq.~(\ref{skkbesigma}) separates the self-energy matrix $\Sigma_{m_1m_2}^{i_1i_2}(t,t')$ into a time-local Hartree-Fock part ($\vec{\Sigma}^\mathrm{HF}$) and a contribution $\vec{\Sigma}^\mathrm{corr}$ that accounts for electron-electron ($e$-$e$) correlation and memory effects. However, as an exact treatment of $e$-$e$ correlations is impractical, we have to apply many-body approximations for which the second Born diagrams provides one of the most basic models. Hence, besides the general form of the HF self-energy
\begin{eqnarray}
\label{sigmahf}
 \Sigma^{\mathrm{HF},i_1i_2}_{m_1m_2}(t)&=&-\mathrm{i}\,\left\{\sigma\,\delta^{i_1i_2}_{m_1m_2}\sum_{i_3m_3}g_{m_3m_3}^{i_3i_3}(t,t^+)\,\tilde{u}_{m_1m_3}^{i_1i_3}\,-\,g_{m_2m_1}^{i_2i_1}(t,t^+)\,\tilde{u}_{m_2m_1}^{i_2i_1}\right\}\;,
\end{eqnarray}
where $t^+$ denotes $t\rightarrow t+\epsilon_{\geq0}$ we, in second Born approximation, have
\begin{eqnarray}
\label{sigma2ndb}
 \Sigma^{\mathrm{corr},i_1i_2}_{m_1m_2}(t,t')&=&\sum_{i_3m_3}\sum_{i_4m_4}\left\{\sigma\,g^{i_1i_2}_{m_1m_2}(t,t')\,g^{i_3i_4}_{m_3m_4}(t,t')\,-\,g^{i_1i_4}_{m_1m_4}(t,t')\,g^{i_3i_2}_{m_3m_2}(t,t')\right\}\,\times\nonumber\\
&&\hspace{4pc}\times\;\;g^{i_4i_3}_{m_4m_3}(t',t)\,\tilde{u}_{m_1m_4}^{i_1i_4}\,\tilde{u}_{m_2m_3}^{i_2i_3}\;.
\end{eqnarray}
Eqs.~(\ref{sigmahf}) and (\ref{sigma2ndb}) involve the spin degeneracy factor $\sigma\in\{1,2\}$, the matrix elements $\tilde{u}_{m_1m_2}^{i_1i_2}$ of Eq.~(\ref{umat1}) and show a very simple form compared to the situation when a general basis is applied, e.g.~\cite{balzer09_prb}. The reason for this is the subtle structure of the FE-DVR. In detail, the Hartree term is completely diagonal [including a single sum over $n_b$ elements] and the exchange term involves only a product of two matrix elements. For the second-order Born terms, the degree of simplification is even more drastic: In a general basis representation, two sums are required for each full vertex point in the second-order self-energy diagrams and, additionally, a single sum is needed for the start- and the end-point. This leads to a scaling with ${\cal O}(n_b^6)$. However, in our case, due to the diagonality of the two-electron integrals, cf.~Eq.~(\ref{umat1}), the evaluation reduces remarkably to a scaling with ${\cal O}(n_b^2)$ per matrix element.

For the atomic and molecular model calculations to be outlined as first benchmarks of the method in Sec.~\ref{sec:ammodels}, we, in this contribution, restrict ourselves to the (equilibrium) ground-state properties. For completeness, we give a short summary of the main computational steps. First, in the FE-DVR picture, the Hartree-Fock equilibrium Green's function, denoted $\vec{g}^0(\tau)$, follows from
\begin{eqnarray}
\label{hamhf}
 h_{m_1m_2}^{0,i_1i_2}[\vec{g}^0(0^-)]=t_{m_1m_2}^{i_1i_2}+v_{m_1m_2}^{i_1i_2}+\Sigma^{0,i_1i_2}_{m_1m_2}[\vec{g}^0(0^-)]\;,
\end{eqnarray}
where $g^{0,i_1i_2}_{m_1m_2}(\tau)$ is the HF approximation of $g^{i_1i_2}_{m_1m_2}(t,t')|_{t-t'=0+\mathrm{i}\tau}$ with $\tau\in[-\beta,0]$ and $\beta$ is the inverse temperature, and $\Sigma^0$ is defined via Eq.~(\ref{sigmahf}), but with $\vec{g}$ being replaced by matrix $\vec{g}^0$. Thereby, Eq.~(\ref{hamhf}) has to be solved to self-consistency by iteration and as result we obtain
\begin{eqnarray}
\label{hfgf}
 g^{0,i_1i_2}_{m_1m_2}(\tau)=\sum_{i m}c_{mm_1}^{ii_1}f_\beta(\epsilon_m^i-\mu)\,e^{-\tau(\epsilon_m^i-\mu)}c_{mm_2}^{ii_2}\;.
\end{eqnarray}
Here, the vector $\epsilon_m^i$ contains the energy eigenvalues of $\vec{h}^0$, the matrix $c_{mm'}^{ii'}$ summarizes the corresponding eigenvectors, and the chemical potential $\mu$ is determined by normalization of the Fermi distribution: $N=\sum_{i m} f_\beta(\epsilon_m^i-\mu)$.
Eq.~(\ref{hfgf}) solves a simple differential equation, see e.g.~\cite{balzer09_prb}, and corrections due to $e$-$e$ correlations in second Born approximation are obtained by insertion into the Dyson equation~~\cite{dahlen05} (the SKKBE in the limit $t-t'=0+\mathrm{i}\tau$) for the full Matsubara Green's function $\vec{g}^M(\tau)$:
\begin{eqnarray}
\label{deq}
g^{M,i_1i_2}_{m_1m_2}(\tau)&=&g^{0,i_1i_2}_{m_1m_2}(\tau)\,+\,I^{(2)\,i_1i_2}_{m_1m_2}(\tau)\;,\\
I^{(2)\,i_1i_2}_{m_1m_2}(\tau)&=&\sum_{im}
\int_0^\beta\!\dint{}{\bar{\tau}}\,g^{0,i_1,i}_{m_1m}(\tau-\bar{\tau})\,I^{(1)\,ii_2}_{mm_2}(\bar{\tau})\;,\nonumber\\
I^{(1)\,i_1i_2}_{m_1m_2}(\tau)&=&
\sum_{im}\int_0^\beta\!\dint{}{\bar{\tau}}\left\{\Sigma^{M,i_1i}_{m_1m}(\tau-\bar{\tau})\,-\,\delta(\tau-\bar{\tau})\,\Sigma^{0,i_1i}_{m_1m}\right\}g^{M,ii_2}_{mm_2}(\bar{\tau})\;,\nonumber
\end{eqnarray}
where $\vec{\Sigma}^M(\tau)=\delta(\tau)\vec{\Sigma}^\mathrm{HF}(\tau)+\vec{\Sigma}^\mathrm{corr}(\tau)$ with $\vec{g}^M$ instead of $\vec{g}$ in Eqs.~(\ref{sigmahf}) and (\ref{sigma2ndb}). The convolution integrals in Eq,~(\ref{deq}) are performed in sequence by direct integration. This has been found to be more stable and controllable than the method applied before in Refs.~\cite{dahlen07_prl,balzer09_prb}. Once the self-consistent $\vec{g}^M(\tau)$ is computed, we have direct access to many observables, e.g.~the one-electron density is obtained as $n(x)=\sum_{i_1m_1}\sum_{i_2m_2}\chi_{m_1}^{i_1}(x)\chi_{m_2}^{i_2}(x)g_{m_1m_2}^{M,i_1i_2}(0^-)$. The total energy is computed similar as in Refs.~\cite{dahlen05,balzer09_prb}. Overall, in order to ensure the atoms and molecules being in the ground state, we set the inverse temperature $\beta=100$.

\section{\label{sec:ammodels}Performance for atomic and molecular models}
As first benchmarks and preparatory work for the investigation of the temporal evolution of small atoms and/or molecules following an external (e.g.~laser-induced) perturbation, we here consider their equilibrium-(initial-)state preparation within the FE-DVR context. As examples, we focus on the He atom and the linear molecular ion H$_3^+$, modeled in one spatial dimension. For both two-electron systems, the Coulomb potential is considered in the regularized form $u(|x-x'|)=[(x-x')^2+1]^{-1/2}$ which, from the physical point of view, allows for a transverse extension of the few-particle wave function. Furthermore, $e$-$e$ correlations are treated in second Born approximation.

\begin{figure}
\includegraphics[width=0.625\textwidth]{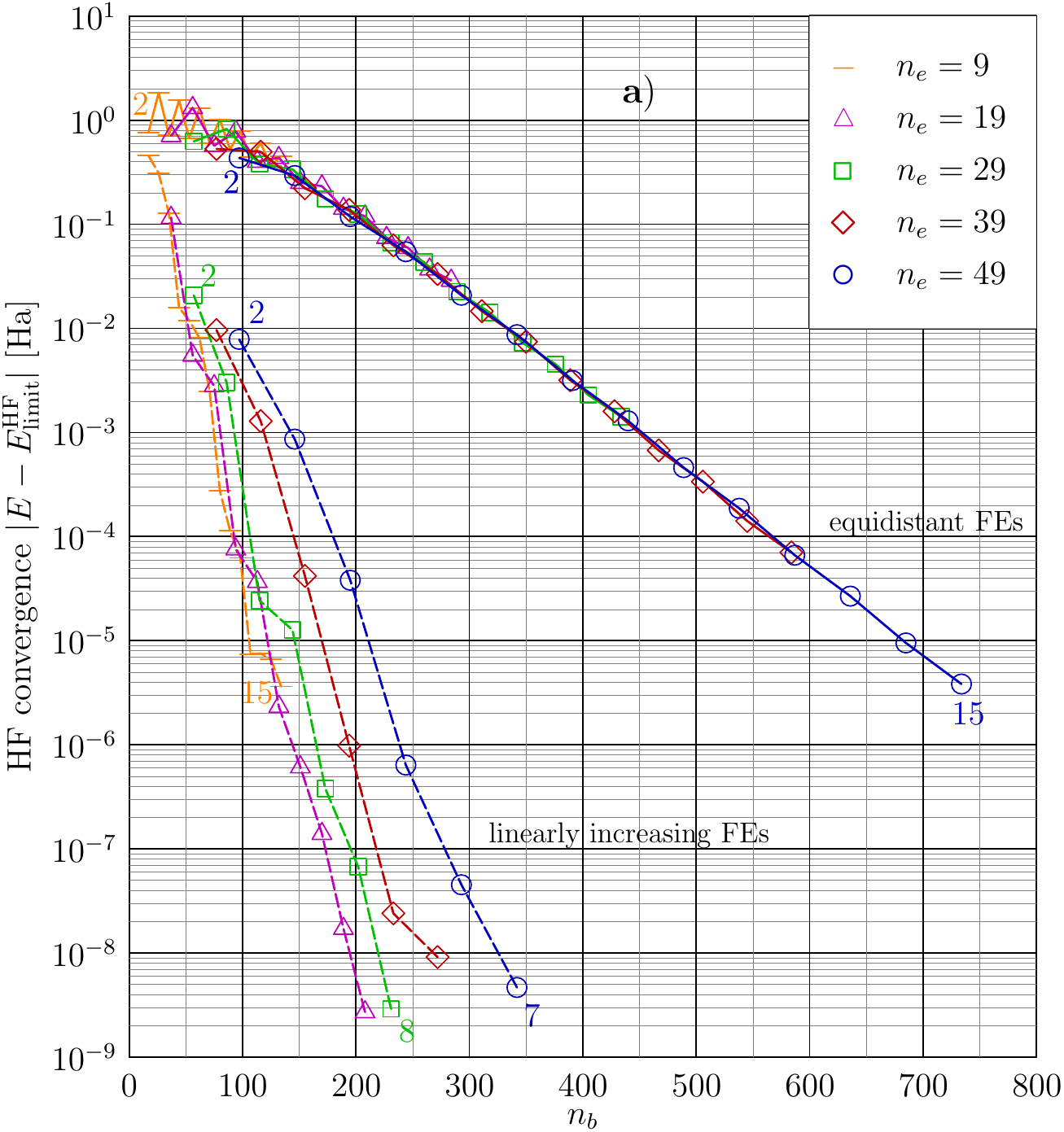}
\includegraphics[width=0.365\textwidth]{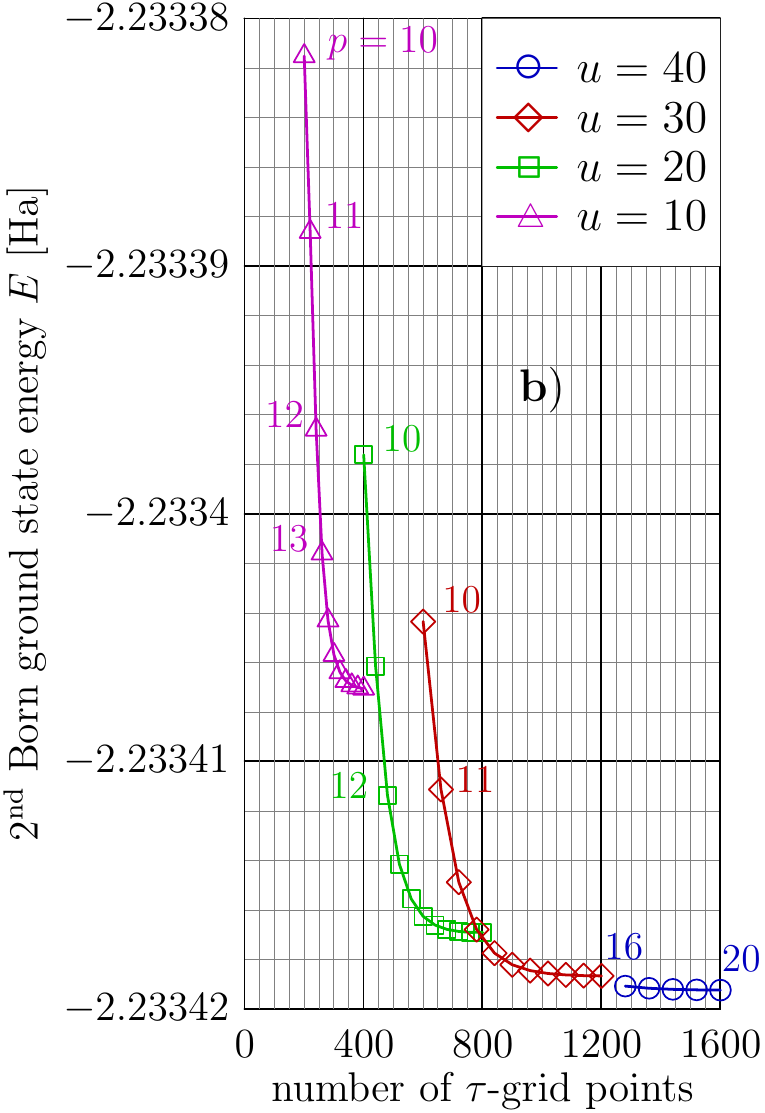}
 \caption{\textbf{a})~Convergence of the Hartree-Fock ground-state energy for the 1D helium atom against basis dimension $n_b$. As indicated, each curve corresponds to a different arrangement of the FEs within a total interval of $200$~a.u.~length. For the upper set of lines the FEs are equally distributed, whereas for the lower set the central FE is one atomic unit wide, and the width of neighboring elements is linearly increasing towards the boundaries. In addition, the number of local DVR basis functions has been varied between $n_g=2$ and $15$, see numbers on the curves. \textbf{b})~He ground-state energy in second Born approximation ($n_b=202$) with respect to different $\tau$-grid parameters $u$ and $p$.
}\label{fig2}
\end{figure}

\subsection{\label{subsec:he}The 1D helium atom}
The helium atom is the most elementary closed-shell two-electron system. In one spatial dimension, it is well modeled by the nucleus potential
$v(x)=-Z\,[{(x-x_0/2)^2+\rho}]^{-1/2}$, where the atomic number is $Z=2$, $x\in[0,x_0]$, and $\rho$ is a regularization parameter. In this setup, the 1D helium atom serves as a fundamental 'testing ground' for multi-electron calculations~\cite{zanghellini04,ruggenthaler09,david_pngf4} and provides many features of the single- and double-ionization dynamics including the so-called knee structure~\cite{dahlen01,lein00}. Considering the singlet state, we refine the model and set $\sigma=2$ in Eqs.~(\ref{sigmahf}) and (\ref{sigma2ndb}), and used $\rho=1$.

Further, for the equilibrium calculations, we have used a FE-DVR basis that covers a total interval of $200$~a.u.~length. This corresponds to a domain that is about $100$ times larger than the characteristic extension of the ground-state wave function or density, cf.~Ref.~\cite{balzer09_pra}. Such a grid extension is more than adequate to resolve the ground-state features of the model (to be discussed here) but will become crucial, when the helium atom is perturbed by external fields and, in turn, electrons start to occupy highly excited or continuum states. In this sense, our results are benchmarks also with relevance for the computation of the system's temporal evolution. In particular, a grid with an extension of about $200$~a.u.~should be well capable to resolve the two-electron resonance states embedded within the one-electron continuum of dipole spectra. This follows from the performance of the few-particle time-dependent Schr\"odinger equation (TDSE) using absorbing potentials, e.g.~\cite{bauch08} and references therein. We note, that also in the FE-DVR approach, such imaginary one-electron potentials that damp reflections at the interval boundaries are easily implemented, just allowing the matrix elements $v_{m_1m_2}^{i_1i_2}$ in Eq.~(\ref{vmat}) to be complex.

The explicit partitioning of the interval into finite elements has been organized as follows: (case I) the interval is divided into equidistant segments, (case II) the central FE has a width of one atomic unit and the width of the surrounding elements is linearly increasing towards the interval boundaries. The effect of these segmentations on the Hartree-Fock ground-state energy convergence of the helium atom is displayed in Fig.~\ref{fig2}~\textbf{a}) as function of the total basis size. In case~I, for equidistant FEs, the convergence is relatively slow with $n_b$ and strongly depends on the number of elements as well as on the number of local DVR basis functions used (see the different symbols and lines). Consequently, more than $n_b>550$ functions are needed for the HF energy to deviate by less than $10^{-4}$~Ha from the converged HF result $E^\mathrm{HF}_\mathrm{limit}=-2.2242096$~Ha. In case~II, the situation is completely different as essentially more basis functions are available in the center region of the interval. This enables a superior representation of the Matsubara Green's function $G^M(x,x';\tau)$ in coordinate space, and leads to adequate convergence at $200$ to $300$ FE-DVR basis functions with an error reduced by several orders of magnitude compared to case~I. Furthermore, the ground-state energy depends less on the number of elements. For completeness, we note that, besides the total energy, an additional indicator for the basis quality, is to look at how well the potential $v(x)$ can be expanded into the chosen FE-DVR basis.

For case~II with $n_e=29$ and $n_g=7$, we have computed the ground-state energy in second Born approximation. Thereby, all quantities are found to be well converged with respect to the basis size, compare with the HF case in Fig.~\ref{fig2}~\textbf{a}). The time-argument $\tau$ in the Matsubara Green's function ($\tau\in[-\beta,0]$) has been discretized using a uniform power mesh~\cite{ku02} with parameters $u$ and $p$---for definition see e.g.~Refs.~\cite{balzer09_prb,dahlen05}. Fig.~\ref{fig2}~\textbf{b}) indicates the convergence with respect to these parameters, where the total number of $\tau$-grid points is given by $2up+1$. At $\beta=100$, a mesh parameter $p\geq10$ ensures the particle number $N=\sum_{im}g^{M,ii}_{mm}(0^-)$ being sufficiently stable during iteration of the Dyson equation~(\ref{deq}). Particularly, with more than $1000$ grid points it is possible to compute the ground-state energy to relatively high precision, $E^\mathrm{2ndB}=-2.233419$~Ha. As result, the electron-electron correlations lower the total energy accounting for $66$\% of the correlation energy and, hence (improving the HF result), approaches the exact ground-state energy which is $-2.2382578$~Ha. For the discussion of other observables such as the one-electron density in HF and second Born approximation, the reader is referred to Ref.~\cite{balzer09_pra}.

\subsection{\label{subsec:h3+}The linear molecule H$_3^+$}
With more than two nuclei, the molecular ion H$_3^+$ can, in 1D, only be realized in its linear version~\cite{alexander09}, where the H-H bonds are oriented parallel to each other. Hence, the electrons move along the molecular axis and the one-electron potential is modeled as
\begin{eqnarray}
\label{h3pv}
 v_d(x)=\left[\left(x-(x_0-d)/2\right)^2+1\right]^{-1/2}\,+\,\left[\left(x-(x_0+d)/2\right)^2+1\right]^{-1/2}\,+\,[x^2+1]^{-1/2}\,+\,\frac{5}{d}\;,
\end{eqnarray}
where $d/2$ denotes the interatomic distance, and the last term (offset) collects all internuclei interaction energy contributions. For the ground-state NEGF calculations, the coordinate space has been constrained to an interval of $50$~a.u.~with a grid of $n_e=13$ FE being linearly increasing, starting from a one atomic unit wide central element. In total, $n_b=142$ FE-DVR basis functions have been used.

\begin{figure}[t]
 \includegraphics[width=\textwidth]{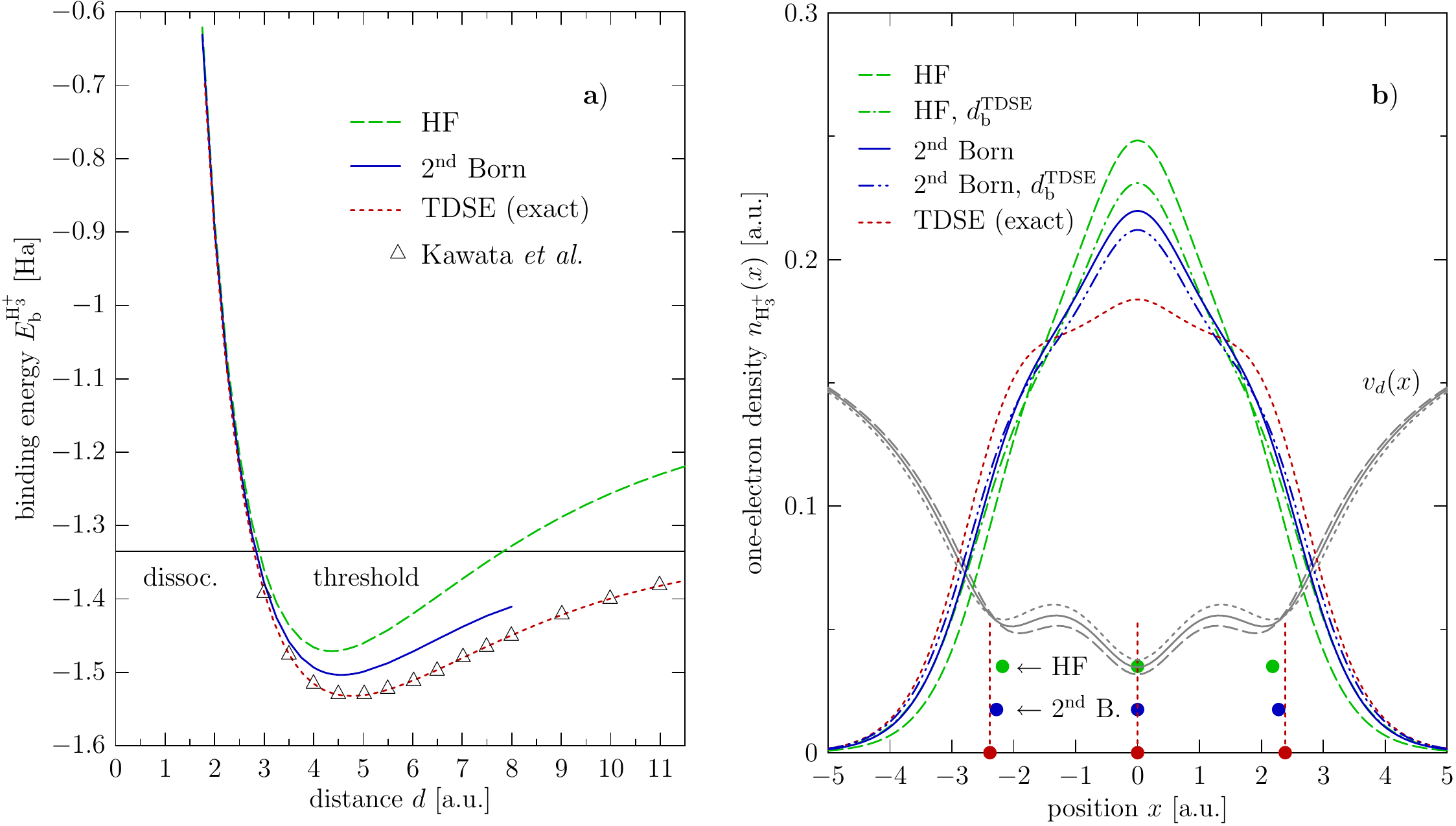}
\caption{\textbf{a})~Binding-energy curves for the linear H$_3^+$ molecule in one spatial dimension for different approximations. The dotted line corresponds to the exact solution obtained from the few-particle time-dependent Schr\"odinger equation (TDSE). The triangles denote the binding energy of Ref.~\cite{kawata01}. \textbf{b})~One-electron density at the equilibrium bond-lengths $2 d_\mathrm{b}$: $4.3654$ (HF),  $4.5579$ (second Born), and $4.7698$ (exact). The dots indicate the self-consistent positions of the nuclei. The gray curves indicate the corresponding potential $v_d(x)$ [scaled by $0.35$ and shifted].}\label{fig3}
\end{figure}

The total binding energy $E_{\mathrm{b}}$ of the singlet state is shown in Fig.~\ref{fig3}~\textbf{a}) against distance $d$ for the exact solution of the TDSE (dotted [and triangles]), the Hartree-Fock (dashed) and the second Born approximation (solid). Over a broad range of internuclear distances, the second Born approximation, thereby, accounts for about $60$-$70$\% of the correlation energy, and---leading overall to a larger bond---essentially improves the HF result.

Furthermore, with a value of $-1.3396$~Ha, H$_3^+$ has the same dissociation threshold as the 1D hydrogen molecule~\cite{balzer09_pra}, but, in addition, leaves behind a positively charged hydrogen ion. However, we note, that the Hartree-Fock and the second Born approximation, cannot resolve this threshold. The reason is that the H$_3^+$ molecule dissociates into open-shell fragments---two hydrogen atoms and a single hydrogen ion. These cannot be represented within a spin-restricted calculation (with $\sigma=2$), and, hence, lead to a strong deviation of $E_\mathrm{b}$ in the limit of large $d$. Regardless of this failure of the ansatz, the NEGF calculations are well capable to describe the behavior of $E_\mathrm{b}$ around the minima in the binding energy curves. Also, the equilibrium positions are consistently reproduced, and the correct trend is observed when $e$-$e$ correlations are being included: the bond-length $d_\mathrm{b}$ shifts to larger nuclear separations, for the specific values obtained see caption of Fig.~\ref{fig3}.

The electron ground-state density of the  molecular ion is displayed in Fig.~\ref{fig3}~\textbf{b}) together with schematic curves for the spatial potentials $v_d(x)$, where $d$ is twice the equilibrium internuclear distance, cf.~Eq.~(\ref{h3pv}). In HF approximation (dashed curve), the density shows a pronounced maximum in the central region of the molecule, whereas the exact density (dotted curve) is essentially less peaked. However, also the exact result does not indicate onset of electron localization, i.e.~does not show separated maxima in $n_{\mathrm{H}^+_3}(x)$. In second Born approximation, corresponding to a lower total energy (cf.~Fig.~\ref{fig3}~\textbf{a})), we resolve the correct trend of this density reduction. Moreover, replacing the self-consistent bond-lengths by the corresponding length obtained from the TDSE (dash-dotted curves) only slightly improves the results for the Hartree-Fock and the second Born approximation. This explains that the substantial differences in the density profiles are unambiguously due to $e$-$e$ correlation effects.

\section{Conclusion}
The FE-DVR ansatz (\ref{gf}) provides an elegant and very efficient way to treat binary interactions in NEGF calculations for inhomogeneous quantum systems. To this end, the method uses a flexible combination of grid (FE) and basis (DVR) strategies, which allows for simple, (semi)-analytical formulas for the matrix elements of the kinetic-, potential- and, especially, the interaction-energy operator, cf.~Sec.~\ref{subsec:matrixelements}. Further, for the most basic model that accounts for particle-particle correlations---the second Born approximation---the use of a FE-DVR basis enables remarkable scaling properties: Only ${\cal O}(n_b^2)$ summations are required for the computation of a single matrix element of the second-order self-energy instead of ${\cal O}(n_b^6)$ summations that are needed in a general basis approach.

Also, we emphasize that the FE-DVR space can, e.g. via finite-element variations, be well adjusted to the problem considered and, thus, allows for an efficient expansion of the NEGF [also for spatially extended hamiltonians] and quick convergence of the observables of interest. This has been exemplified for the He atom and the triatomic molecule H$_3^+$ in Sec.~\ref{subsec:he} and \ref{subsec:h3+}. In addition, the method is found to be stable also for large grids [large basis sets with $n_b>500$] and also for larger particle numbers $N\leq20$, considering interacting fermions in a harmonic trap potential.

Finally, we believe that the FE-DVR method is attractive also for other classes of many-body approximations, such as $GW$ or $T$-matrix calculations, as it will likewise simplify the computation of self-energy contributions of higher than second order. Moreover, though  we, in this contribution, focused on the solution of the Dyson equation, the formalism presented is, in particular, well applicable in nonequilibrium situations solving the full two-time SKKBEs which will be demonstrated in a forthcoming publication.


\begin{thebibliography}{199}
\bibitem{generalization} The generalization to different numbers of basis functions per element is straightforward, and only slightly alters the matrix elements involved.
\bibitem{symmetryrel} The two-electron integrals $u_{ij,\bar{i}\bar{j}}$ are symmetric with respect to interchange of $i\leftrightarrow j$, $\bar{i}\leftrightarrow\bar{j}$ and tupels $(i,j)\leftrightarrow(\bar{i},\bar{j})$.
\end{thebibliography}
\end{document}